\documentclass[twocolumn,secnumarabic,amssymb, nobibnotes, aps, prd,showpacs]{revtex4}
\usepackage{graphicx}
\usepackage{longtable}

\begin{document}

\title{Energy-Momentum Distribution in Static and Non-static Cosmic String Space-times}%

\author{Amir M. Abbassi}
\email{amabasi@khayam.ut.ac.ir}
\author{Saeed Mirshekari}
\email{smirshekari@ut.ac.ir}
 \affiliation{Department of Physics,
University of Tehran, North Kargar Ave, \\Tehran, Iran.}
\author{Amir H. Abbassi} \email{ahabbasi@modares.ac.ir} \affiliation{Department of Physics,
Tarbiat Modares University, P.O.Box 14155-4838, Tehran, Iran}
\date{December 2008}
\begin{abstract}
In this paper, we elaborate the problem of energy-momentum in GR
by energy-momentum prescriptions theory. In this regard, we
calculate M\o ller, Landau-Lifshitz, Papapetrou, Einstein, Bergmann,
Tolman, and Weinberg's energy-momentum complexes in static and
non-static cosmic string space-times. We obtain strong
coincidences between the results. These coincidences can be
considered as an extension of Virbhadra's viewpoint that different
energy-momentum prescriptions may provide some basis to define a
unique quantity. In addition, our results disagree with Lessner's
belief about M\o ller's prescription, and support the Virbhadra's
conclusion about power of Einstein's prescription.

\keywords{Energy-momentum Prescriptions, Cosmic String
space-time.} \pacs{04.20.-q, 04.20.Cv, 98.80.Cq,
04.20.Jb,98.80.Es}
\end{abstract}

\maketitle
 
\section{Introduction}
In classical mechanics and even in special relativity, we can
always introduce a two-indices, symmetric tensorial quantity, i.e.
$T_{a}^{b}$ , which is named as energy-momentum tensor and
represents the energy and momentum of matter and non-gravitational
fields sources. Besides mentioned properties (being tensorial and
symmetric), it has an important special characteristic: it is
localized. This means that in every point of the manifold the
quantity of energy-momentum is conserved. In the other words,
energy-momentum tensor is a divergenceless quantity. In fact, in
any local point of manifold no contribution of this quantity
produces and no eliminates. We have:
\begin{equation}\label{1}
T_{a ,b}^{b}=0
\end{equation}
Eq.(\ref{1}) is definition of
energy-momentum conservation and known as conservation laws. Since
the energy and momentum are two important, conserved quantities in
physics, people are interested to keep it (as usual form of
conservation laws) unchanged in all fields of physics, especially
in general relativity's theory. But in GR, ordinary derivatives
transform to covariant derivatives. So we have~\cite{weinberg}:
\begin{equation}\label{2}
T^{b}_{a
;b}=\frac{1}{\sqrt{-g}}(\sqrt{-g}T^{b}_{a})_{,b}-\Gamma_{a
c}^{b}T_{b}^{c}=0
\end{equation}
It is obvious from Eq.(\ref{2}) that $ T_{a}^{b} $  no longer
satisfies $ T_{a ,b}^{b}=0 $, but as noted before, we are
interested to have a similar equation in GR. We  add an additional
term to $ T_{a}^{b} $, e.g. $ t_{a}^{b} $, so that the summation
of these two terms remains divergenceless. In reality, the
quantity that is actually conserved in the sense of
Eq.(\ref{1}) is some effective quantity
which is given (in one variant) by Eq.(20.18) of MTW [2] as $
_{\textrm{\tiny eff}}T_{a}^{b}=(T_{a}^{b}+ t_{a}^{b}) $. In other
variants, we obtain
\begin{equation} \label{3}
_{\textrm{\tiny
eff}}T^{b}_{a}=(-g)^{\frac{n}{2}}(T^{b}_{a}+t^{b}_{a})
\end{equation}
where $ g=det(g_{a b}) $  and  $ n $ is a positive integer that
indicates the weight. For each of these $ _{\textrm{\tiny
eff}}T^{b}_{a} $, Eq.(\ref{2}) can be rewritten as:
\begin{equation}\label{4}
_{\textrm{\tiny eff}}T^{b}_{a ,b}=0
\end{equation}
Conserved quantity $ _{\textrm{\tiny eff}}T^{b}_{a} $ refers to
the flux and density of energy and momentum of gravitational
systems.
In fact, coming from SR to GR, we add a contribution of
gravitational fields,$ t_{a}^{b} $ to the contribution of matter
and all non-gravitational fields,$ T_{a}^{b} $ . Einstein,
himself, proposed the first prescription for $ _{\textrm{\tiny
eff}}T_{a}^{b} $ just after GR's formulation in 1916. Then, many
other persons such as M\o ller [3], Landau-lifshitz [4], Papapetrou
[5], Bergmann [6], Tolman [7], and Weinberg [1] gave different
prescriptions. All proposed expressions are called energy-momentum
complexes, because they can be expressed as a combination of a
tensor, $ T_{a}^{b} $, and a pseudo-tensor, $ t_{a}^{b} $.
However, by using this (adding $ t_{a}^{b} $ to $ T_{a}^{b} $  )
we could solve the problem of being non-zero divergence of
energy-momentum tensor, but some serious problems arise.
Actually, it can be shown that $ t_{a}^{b} $  does not obey tensor
transformations. This non-tensorial property of $ _{\textrm{\tiny
eff}}T_{a}^{b} $   has caused that these complexes not to satisfy
the required covariance and be coordinate dependent; it is the
main problem of using energy-momentum complexes. Some authors
tried to  introduce new coordinate independent prescriptions. In
fact, except few prescriptions including Penrose [8], M\o ller [3],
and Komar's [9] prescriptions, for other energy-momentum complexes
one gets
physically meaningful results only in Cartesian coordinate system.\\
Next problem is that it is not necessary for $ _{\textrm{\tiny
eff}}T_{a}^{b} $ to be symmetric in all prescriptions. We can
define conserved angular momentum quantity only for symmetric
prescriptions [1]. In this regard, anti-symmetric characteristic
of Einstein's prescription was the main motivation for Landau and
Lifshitz to look for an alternative prescription for
energy-momentum which is symmetric. We have listed some prevalent
and well-known prescriptions (that we have used in the next
sections) and their
properties in table \ref{table-symmetry}.

\begin{table}
\centering \caption{Comparison of the differences between
energy-momentum prescriptions in symmetry and suitable coordinate
systems needed.} \label{table-symmetry}
\begin{tabular}{|c|c|c|}
\hline
Prescription & Coordinate System & Symmetry \\
\hline M\o ller & any coordinate system & anti-symmetric \\
Landau-Lifshitz & cartesian & symmetric \\
Papapetrou & cartesian & symmetric \\
Einstein & cartesian & anti-symmetric \\
Bergmann & cartesian & non-symmetric \\
Tolman & cartesian & non-symmetric \\
Weinberg & cartesian & symmetric \\
\hline
\end{tabular}
\end{table}

For making the subject clearer, it should be noted that $
_{\textrm{\tiny eff}}T_{a}^{b} $  can be written as the divergence
of some \textit{super-potential} $ H_{a}^{[b c]} $  that is
anti-symmetric in its two upper indices [10] as
\begin{equation}
_{\textrm{\tiny eff}}T_{a}^{b}=H_{a ,c}^{[b c]}
\end{equation}
In addition, a new function like $ U_{a}^{b c} $ can also play
the role of $ H_{a}^{[b c]} $ if it has the following conditions:
\begin{equation}
U_{a}^{b c}=H_{a}^{[b c]}+\Psi_{a}^{b c}\;(\Psi_{a ,c}^{b
c}\equiv 0,\: or\: \Psi_{a ,cb}^{b c} )
\end{equation}
Then, the quantity $ \Theta_{a}^{b} $  which is defined by this
new super-potential remains conserved locally:
\begin{equation}
\Theta_{a}^{b}=U_{a ,c}^{b c}\Rightarrow \Theta_{a ,b}^{b}=0
\end{equation}
Using this freedom on the choice of the super-potential, authors
like Einstein and Tolman arrived through different methods at the
following super-potentials [11]:
\begin{equation}
H_{a}^{[b c]}=\frac{1}{2\kappa}\tilde{g}_{a e}(\tilde{g}^{e
b}\tilde{g}^{d c}-\tilde{g}^{e  c}\tilde{g}^{d
b})_{,d}\;\;\;(Einstein)
\end{equation}
\begin{equation}
\tau_{a}^{b c}=H_{a}^{[b
c]}+\frac{1}{2\kappa}(\delta_{a}^{c}\tilde{g}^{d
b}-\delta_{a}^{d}\tilde{g}^{c b})_{,d}\;\;\;(Tolman)
\end{equation}
where $\tilde{g}^{a b}=\sqrt{-g}g^{a b}$.

Considering above discussion, there are many prescriptions for new
energy-momentum density ($ _{\textrm{\tiny eff}}T_{a}^{b} $)
[1,3-9] which their differences are in a curl term. Each of them
has its own advantages and disadvantages and it has not proved any
preferences between them. However, Palmer [12] and Virbhadra [13]
discussed the importance of Einstein's energy-momentum
prescription and Lessner [14] believed that M\o ller's prescription
is a powerful tool for calculating the energy-momentum in GR.

The problems associated with concept of energy-momentum complexes
resulted in some researchers even doubting the concept of
energy-momentum localization. Misner et al. [2] argued that to
look for a local energy-momentum is looking for the right answer
to the wrong question. He showed that the energy can be localized
only in systems which have spherical symmetry. Cooperstock and
Sarracino [15] proved that if energy is localizable for spherical
systems, then it can be localized in any system. In 1990, Bondi
[16] argued that a non-localizable form of energy is not allowed
in GR. Recently, besides energy-momentum prescriptions theory, it
was suggested another solution for energy problem in GR that is in
agreement with energy-momentum prescriptions theory about
localization of energy, i.e. \textit{Tele-Parallel Gravity} (for
example see [17]). On the other hand, some people do not believe
in localization of energy and momentum in GR. In addition, some
physicists propose a new concept in this regard: \textit{quasi-localization}
(for example see [18]). Unlike energy-momentum prescriptions theory,
quasi-localization theory does not restrict one to use particular coordinate system,
but this theory have also its drawbacks [19,20]. In general, there is no generally
accepted definition for energy and momentum in GR until now. Chang et al.
in ref. [21] showed that every energy-momentum complex can be associated
with distinct boundary term which gives the quasi local energy-momentum.
By this way, he dispels doubts expressed about the physical meaning of
energy-momentum complexes.\\
For a long time, there have been an uncertainty that different
energy-momentum complexes would give different results for a given
space-time. Many researches considered different prescriptions and
obtained interesting results. Virbhadra et al. [13,22-25]
investigated several examples of the space-times and showed that
different prescriptions could provide exactly the same results for
a given space-time. Aguirregabiria et al. [26] proved the
consistency of the results obtained by using the different
energy-momentum complexes for any Kerr-Schild class metric and
revived the energy-momentum prescriptions theory after a long
period of time.\\
In this paper we extend the previous works by calculating the
energy of static and non-static cosmic string space-times in a
specific region by seven well-known energy-momentum prescriptions:
M\o ller, Landau-Lifshitz, Papapetrou, Einstein, Bergmann, Tolman,
and Weinberg. We obtain encouraging results which show interesting
coincidences between the results calculated by different
prescriptions. Our results about M\o ller's prescription disagree
with Lessner's viewpoint but support the Virbhadra's conclusion
that Einstein's prescription is the best available method for
computing energy-momentum in a given space-time. The rest of the
paper is organized as follows: in section (2) we introduce
energy-momentum complexes those we use in next sections. Section (3)
contains an introduction to static and non-static cosmic string
space-times. The method and results of calculations are written in
section (4). In section (5) we summarize and conclude with some
remarks and discussions.\\
Conventions: We use geometrized units in which the speed of light
in vacuum $ c $ is
taken to be equal to 1 and the metric has signature $(+ - - -)$ . Latin indices take values 0...3.

\section{Energy-Momentum Prescriptions}
Among many different forms proposed for energy-momentum pseudo-tensors, in this article we shall use M\o ller, Landau-Lifshitz, Papapetrou,
Einstein, Bergmann, Tolman, and Weinberg's prescriptions. Just in the same direction as the previous works in the literature here we try to show the compatibilities and to find out any existing discrepancies between predictions of these prescriptions when applying to static and non-static cosmic string space-times. Specific forms of
each energy-momentum pseudo-tensor, conservation laws, and energy-momentum
4-vectors are listed in Table II briefly. Interested readers can
refer to the mentioned references for details. In the last column of
Table \ref{Energy Prescriptions} Gauss' theorem is used. In the
surface integrals $ n_{a} $ represent the components of a normal one form
over an infinitesimal surface element $ds$. The results of calculations according to each of the individual forms will be shown in the following sections. 

\begin{table*}
  \centering
  \caption{Energy-momentum Prescriptions}\label{Energy Prescriptions}

\begin{ruledtabular}
\begin{tabular}{|c|c|c|c|}
&&&\\
  Prescription & Energy-momentum pseudo-tensor & Conservation & Energy-momentum\\

&& Laws& 4-vector  \\
&&&\\
\hline
&&&\\
  M\o ller [3] & $M_{i}^{k}=\frac{1}{8\pi}\chi_{i ,l}^{k l}$ &$
\frac{\partial M_{i} ^{k} }{\partial x^{k}}=0$ &$
P_{i}=\int\int\int M_{i} ^{0}dx^{1}dx^{2}dx^{3}$\\
 & $\chi_{i}^{k l}=\sqrt{-g}(\frac{\partial g_{ip}}{\partial
x^{q}}-\frac{\partial g_{iq}}{\partial x^{p}})g^{kq}g^{lp}$&
&$=\frac{1}{8\pi}\int\int\chi_{i}^{0 a}n_{a}ds$\\
&&&\\
\hline
&&&\\
  Landau-Lifshitz[4]&$ L^{ik}=\frac{1}{16\pi}\lambda^{iklm}_{,lm}$& $\frac{\partial L^{i k} }{\partial x^{k}}=0
 $& $P^{i}=\int\int\int L^{i 0}dx^{1}dx^{2}dx^{3}$\\
  & $\lambda^{iklm}=-g(g^{ik}g^{lm}-g^{il}g^{km})_{,m}$& &$=\frac{1}{16\pi}\int\int \lambda^{i0\alpha m}_{,m}n_{a}ds$\\
&&&\\
\hline
&&&\\
   & $\Sigma^{ik}=\frac{1}{16\pi}N^{iklm}_{,lm}$&&$ P^{i}=\int\int\int \Sigma^{i 0}dx^{1}dx^{2}dx^{3}$\\

 Papapetrou[5] &$N^{iklm}=\sqrt{-g}(g^{ik}\eta^{lm}-g^{il}\eta^{km}+g^{lm}\eta^{ik}-g^{lk}\eta^{im})$&$\frac{\partial \Sigma^{i k} }{\partial x^{k}}=0$&$=\frac{1}{16\pi}\int\int N^{i0\alpha m}_{,m}n_{a}ds$\\

 &$\eta^{ik}=diag(1,-1,-1,-1)$&&\\
&&&\\
\hline
&&&\\

  Einstein[3] & $\Theta_{i}^{k}=\frac{1}{16\pi}H_{i ,l}^{k l}$
& $\frac{\partial \Theta_{i} ^{k} }{\partial x^{k}}=0$
 & $P_{i}=\int\int\int \Theta_{i} ^{0}dx^{1}dx^{2}dx^{3}$ \\
&$H_{i}^{kl}=-H_{i}^{lk}=\frac{g_{in}}{\sqrt{-g}}[-g(g^{kn}g^{lm}-g^{ln}g^{km})]_{,m}$&&$=\frac{1}{16\pi}\int\int H_{i}^{0 a}n_{a}ds$\\
&&&\\
\hline
&&&\\

   & $B^{ik}=\frac{1}{16\pi}\beta^{ikm}_{,m}$ && $P^{i}=\int\int\int B^{i 0}dx^{1}dx^{2}dx^{3}$\\

Bergman[6]&$\beta^{ikm}=g^{ir}\nu_{r}^{km}$&$\frac{\partial B^{i
k} }{\partial x^{k}}=0$&$=\frac{1}{16\pi}\int\int
\beta^{i0\alpha m}_{,m}n_{a}ds$\\
&$\nu_{i}^{kl}=-\nu_{i}^{lk}=\frac{g_{in}}{\sqrt{-g}}[-g(g^{kn}g^{lm}-g^{ln}g^{km})]_{,m}$&&\\
&&&\\
\hline
&&&\\

 & $T_{i}^{k}=\frac{1}{8\pi}U_{i ,l}^{k l}$
 & & $P_{i}=\int\int\int T_{i} ^{0}dx^{1}dx^{2}dx^{3}$\\

Tolman[7]&$U_{i}^{k
l}=\sqrt{-g}(-g^{pk}V_{ip}^{l}+\frac{1}{2}g_{i}^{k}g^{pm}V_{pm}^{l})$&$\frac{\partial
 T_{i} ^{k} }{\partial x^{k}}=0$&$=\frac{1}{8\pi}\int\int U_{i}^{0 a}n_{a}ds$\\
&$V_{jk}^{i}=-\Gamma_{jk}^{i}+\frac{1}{2}g^{i}_{j}\Gamma_{mk}^{m}+\frac{1}{2}g^{i}_{k}\Gamma_{mj}^{m}$&&\\
&&&\\
\hline
&&&\\

 &$W^{ik}=\frac{1}{16\pi}D^{i k l}_{,l}$&&$P^{i}=\int\int\int
W^{i0}dx^{1}dx^{2}dx^{3}$
\\

Weinberg[1]&$D^{ijk}=\frac{\partial{h_{a}^{a}}}{\partial{x_{i}}}\eta^{jk}-
\frac{\partial{h_{a}^{a}}}{\partial{x_{j}}}\eta^{ik}-\frac{\partial{h^{ai}}}{\partial{x^{a}}}\eta^{jk}+\frac{\partial{h^{aj}}}{\partial{x^{a}}}\eta^{ik}+\frac{\partial{h^{ik}}}{\partial{x_{j}}}-\frac{\partial{h^{jk}}}{\partial{x_{i}}}$&$\frac{\partial
W ^{ik} }{\partial x^{k}}=0$&$=\frac{1}{16\pi}\int\int
D^{i0a}n_{a}ds
$\\
&$h_{ik}=g_{ik}-\eta_{ik}$&&\\
&&&\\
\end{tabular}
\end{ruledtabular}
\end{table*}

\section{Cosmic String Space-times}

Very early universe is one of the hot and interesting subjects of
theoretical physics that its structure has remained as a mystery
until today. Cosmologists are generally assumed that at very early
stages of its evolution, the universe has gone through a number of
phase transitions. One of the immediate consequences of this phase
transitions is the formation of defects or mismatches in the
orientation of the Higgs field in causally disconnected regions
[27]. Cosmic strings are one of remarkable topological defects
that have received particular attention because of their
cosmological implications. The double quasar problem can be
explained by strings and galaxy formation might also be generated
by density fluctuation in the early universe due to strings [28].

Suppose an infinitely long, thin, straight, static string lying
along $z$ axis with the following stress-energy tensor
\begin{equation}
T_{a}^{b}=\mu \delta(x) \delta(y) diag(1,0,0,1)
\end{equation}
where $\mu$ is the mass per unit length of the string in the $z$
direction. Considering space-time symmetries, Einstein's field
equations lead to well-known solution for case $\Lambda=0$ in
polar cylindrical coordinate system $ (t,\rho,\phi,z) $ [29-31]:
\begin{equation}\label{metric static=0}
ds^{2}=dt^{2}-dz^{2}-d\rho^{2}-(1-4G\mu)^{2}\rho^{2}d\phi^{2}
\end{equation}
For $\Lambda\neq0$ Einstein's field equations lead to general form
of static cosmic string space-time with the following line element
in polar cylindrical coordinate system $ (t,\rho,\phi,z) $ [32]:
\begin{eqnarray}\label{metric static =/ 0}
ds^{2}&=&\cos^{\frac{4}{3}}(\frac{\sqrt{3\Lambda}}{2}\rho)(dt^{2}-dz^{2})-d\rho^{2}\\
&&-\frac{4(1-4G\mu)}{3\Lambda}\cos^{\frac{4}{3}}(\frac{\sqrt{3\Lambda}}{2}\rho)\times
\tan^{2}(\frac{\sqrt{3\Lambda}}{2}\rho)d\phi^{2} \nonumber
\end{eqnarray}
where for $\Lambda\rightarrow 0$ reduces to the previous metric,
Eq.(\ref{metric static=0}). Investigating the non-static solution
of the cosmic strings Einstein's field equations lead to
non-static cosmic string space-time with the following line
element in polar cylindrical coordinate system $ (t,\rho,\phi,z) $
[32]:
\begin{equation}\label{metric non static}
ds^{2}=dt^{2}-e^{2\sqrt{\frac{\Lambda}{3}}t}[d\rho^{2}+(1-4G\mu)^{2}\rho^{2}d\phi^{2}+dz^{2}]
\end{equation}
where for $\Lambda\rightarrow0$ or $t=0$ reduces to
Eq.(\ref{metric static=0}). In the next section we calculate the
energy of these space-times by different energy-momentum
prescriptions.

\section{Calculations}
\subsection{Method}
As mentioned in section (2), for calculating the energy of a given
space-time in a specific region by energy-momentum prescriptions
we should integrate energy-momentum super-potentials over a
suitable surface in space-time. So, we should calculate the
super-potential components, and then indicate normal vector over
infinitesimal surface element . In the two next subsections
integrations are over a cylindrical surface surrounding the length
$L$ from the string symmetrically with radius $\rho$.

It should be noted that in Cartesian coordinate system
(considering $\phi= \arctan(\frac{y}{x})$, and
$\rho=\sqrt{x^{2}+y^{2}}$) Eq.(\ref{metric static =/ 0}) and
Eq.(\ref{metric non static}) transform to the following line
elements (Eq.(\ref{metric car static =/ 0}),(\ref{metric car non
static})) respectively:
\begin{eqnarray}\label{metric car static =/ 0}
ds^{2}&=&\cos^{\frac{4}{3}}\alpha dt^{2}\\
&&-\frac{1}{3}\frac{3\Lambda x^{2}
(x^{2}+y^{2})\cos^{\frac{2}{3}}\alpha
+4a^{2}y^{2}\sin^{2}\alpha}{\Lambda(x^{2}+y^{2})\cos^{\frac{2}{3}}\alpha}dx^{2}\nonumber\\
&&-\frac{2}{3}\frac{3\Lambda(x^{2}+y^{2})\cos^{\frac{2}{3}}\alpha-4a^{2}\sin^{2}\alpha}{\Lambda(x^{2}+y^{2})^{2}\cos^{\frac{2}{3}}\alpha}xy
dxdy\linebreak\nonumber\\
&&-\frac{1}{3}\frac{3\Lambda x^{2}
(x^{2}+y^{2})\cos^{\frac{2}{3}}\alpha
+4a^{2}x^{2}\sin^{2}\alpha}{\Lambda(x^{2}+y^{2})\cos^{\frac{2}{3}}\alpha}dy^{2}\nonumber\\
&&-\cos^{\frac{4}{3}}\alpha dz^{2}\nonumber
\end{eqnarray}
where $ a=(1-4G\mu) $ and $ \alpha=\frac{\sqrt{3\Lambda}}{2}
\rho$.
\begin{eqnarray}\label{metric car non static}
ds^{2}&=&dt^{2}-e^{\alpha
t}\frac{x^{2}+a^{2}y^{2}}{x^{2}+y^{2}}dx^{2}\\
&&+2e^{\alpha t}\frac{a^{2}-1}{x^{2}+y^{2}}x y dx dy-e^{\alpha
t}\frac{y^{2}+a^{2}x^{2}}{x^{2}+y^{2}}dy^{2}-e^{\alpha
t}dz^{2}\nonumber
\end{eqnarray}
where $ a=(1-4G\mu) $ and $ \alpha=2\sqrt{\frac{\Lambda}{3}}$
(remember that $\alpha$ in Eq.(\ref{metric car static =/ 0}) is
different from that is defined in Eq.(\ref{metric car non
static})).

Everywhere we use $ds=\rho\: d\phi\: dz$ as infinitesimal surface
element. In polar cylindrical coordinate system (allowed in
M\o ller's Prescription) we have $ n_{a}=(0,1,0,0) $, and in
Cartesian coordinate system (all prescriptions) we have $
n_{a}=(0,\frac{x}{\rho},\frac{y}{\rho},0) $ . Summarize all above,
for calculating the energy, after extracting needed
super-potential components, we must calculate surface integrals
over a cylindrical surface with suitable normal vector $ n_{a} $
that depends on used coordinate system. Following this method, in
the next subsections we bring needed non-zero components of
super-potentials, and final energy results (i.e. $ P^{0} $ or $
P_{0} $ ) which are calculated by different energy-momentum
prescriptions. Exact expressions of energy (except for M\o ller
prescription) are very well-defined but long and complicated. So,
we restricted ourselves to study the manner of energy around
$\Lambda=0$. Calculations for static($\Lambda\neq0$) and
non-static cosmic string space-times are classified in two
separate subsections. Meanwhile, it should be noticed that we have
done similar calculations by using static($\Lambda=0$) cosmic
string space-time (line element Eq.(\ref{metric static=0})) that
their results are presented in Table III directly.

\subsection{Static Cosmic String ($\Lambda\neq 0$)}
Defining  $ \alpha=\frac{\sqrt{3\Lambda}}{2}\rho $  and  $
a=1-4G\mu $ energy can be calculated by different energy-momentum
prescriptions as follows:

\subsubsection{M\o ller prescription}
Using Table II and Eq.(\ref{metric static =/ 0}) we find that
non-zero needed components of $ \chi_{i}^{kl} $  are:
\begin{equation}
\chi_{t}^{tx}=-\frac{4}{3}\frac{a\sin^{2}\alpha}{x^{2}+y^{2}}x
\end{equation}
\begin{equation}
\chi_{t}^{ty}=-\frac{4}{3}\frac{a\sin^{2}\alpha}{x^{2}+y^{2}}y
\end{equation}
Using surface integral (Table II) energy can be obtained as
\begin{equation}\label{18}
_{M}E=-\frac{1}{3}a\,L\sin^{2}\alpha
\end{equation}
after Taylor expansion around $ \Lambda=0 $ we have:
\begin{equation}
_{M}E=-\frac{1}{4}a\rho^{2}L\Lambda+\frac{1}{16}a\rho^{4}L\Lambda^{2}-\frac{1}{160}a\rho^{6}L\Lambda^{3}+...
\end{equation}
That for $ \Lambda=0 $ vanishes immediately. In addition, as we
expect, same energy expression (Eq.(\ref{18})) are obtained by
using M\o ller energy-momentum prescription with this metric in
polar cylindrical coordinate system instead of Cartesian
coordinate system.

\subsubsection{Landau-Lifshitz prescription}
Using Table II and Eq.(\ref{metric car static =/ 0}) we find that
non-zero needed components of $ \lambda^{iklm} $ are:
\begin{equation}
\lambda^{ttxx}=-\frac{1}{3}\cos^{\frac{2}{3}}\alpha\frac{3y^{2}\Lambda\cos^{\frac{2}{3}}\alpha
(x^2+y^2)+4a^2x^2\sin^2\alpha}{\Lambda(x^2+y^2)^2}
\end{equation}
\begin{equation}
\lambda^{ttyy}=-\frac{1}{3}\cos^{\frac{2}{3}}\alpha\frac{3x^{2}\Lambda\cos^{\frac{2}{3}}\alpha
(x^2+y^2)+4a^2y^2\sin^2\alpha}{\Lambda(x^2+y^2)^2}
\end{equation}
\begin{equation}
\lambda^{ttzz}=-\frac{4}{3}\frac{a^2\sin^2\alpha}{\Lambda(x^2+y^2)\cos^{\frac{2}{3}}\alpha}
\end{equation}
\begin{equation}
\lambda^{ttxy}=\lambda^{ttyx}=\frac{1}{3}\cos^{\frac{2}{3}}\alpha\frac{3\Lambda\cos^{\frac{2}{3}}\alpha
(x^2+y^2)-4a^2\sin^2\alpha}{\Lambda(x^2+y^2)^2}
\end{equation}
After surface integration (Table II) and Taylor expansion around $
\Lambda=0 $ we obtain:
\begin{equation}
_{LL}E=\frac{(1-a^2)}{8}L+\frac{(3a^2-1)}{16}L\rho^{2}\Lambda-\frac{(7a^2-1)}{128}L\rho^4\Lambda^2+...
\end{equation}

\subsubsection{Papapetrou prescription}
Using Table II and Eq.(\ref{metric car static =/ 0}) we find that
non-zero needed components of $ N^{iklm} $ are:
\begin{eqnarray}
N^{ttxx}&=&-\frac{1}{6}\sqrt{\frac{3}{\Lambda}}\frac{1}{a\cos\alpha\sin\alpha(x^2+y^2)^\frac{3}{2}}\\
&&\times [\cos^\frac{8}{3}(x^2+y^2)(3\Lambda
y^2-4a^2)\nonumber\\
&&+4a^2x^2\sin^2\alpha\cos^2\alpha
+4a^2\cos^\frac{2}{3}\alpha(x^2+y^2)]\nonumber
\end{eqnarray}
\begin{eqnarray}
N^{ttyy}&=&-\frac{1}{6}\sqrt{\frac{3}{\Lambda}}\frac{1}{a\cos\alpha\sin\alpha(x^2+y^2)^\frac{3}{2}}\\
&& \times [\cos^\frac{8}{3}(x^2+y^2)(3\Lambda
x^2-4a^2)\nonumber\\
&&+4a^2y^2\sin^2\alpha\cos^2\alpha
+4a^2\cos^\frac{2}{3}\alpha(x^2+y^2)]\nonumber
\end{eqnarray}

\begin{equation}
N^{ttzz}=-\frac{4\sqrt{3}}{3}\frac{a\sin\alpha}{\sqrt{\Lambda
(x^2+y^2)}\cos^{\frac{1}{3}}\alpha}
\end{equation}
\begin{eqnarray}
N^{ttxy}&=&N^{ttyx}=\\
&&-\frac{\sqrt{3}}{6}\frac{\cos\alpha}{\sqrt{\Lambda(x^2+y^2)}}\nonumber\\
&&\times\frac{3\Lambda\cos^{\frac{2}{3}}\alpha(x^2+y^2)-4a^2\sin^2\alpha}{a(x^2+y^2)\sin\alpha}\nonumber
\end{eqnarray}
After surface integration (Table II) and Taylor expansion around $
\Lambda=0 $ we obtain:
\begin{equation}
_{P}E=\frac{(1-a^2)}{8a}L+\frac{(3a^2-1)}{16a}L\rho^{2}\Lambda-\frac{(17a^2-2)}{320a}L\rho^4\Lambda^2+...
\end{equation}

\subsubsection{Einstein prescription}
Using Table II and Eq.(\ref{metric car static =/ 0}), we find that
complicated quantities of $ H_{t}^{tx} $ and $ H_{t}^{ty} $  are
only non-zero needed components of super-potential. After surface
integration (Table II) and Taylor expansion around $\Lambda=0$ we
obtain:
\begin{equation}
_{E}E=\frac{(1-a^2)}{8a}L+\frac{(3a^2-1)}{16a}L\rho^{2}\Lambda-\frac{(17a^2-2)}{320a}L\rho^4\Lambda^2+...
\end{equation}

\subsubsection{Bergmann prescription}
Using Table II, and Eq.(\ref{metric car static =/ 0}), we find
that complicated quantities of $ B^{ttx} $ and $ B^{tty} $  are
only non-zero needed components of super-potential. After surface
integration (Table II) and Taylor expansion around $\Lambda=0$ we
obtain:
\begin{equation}
_{B}E=\frac{(1-a^2)}{8a}L+\frac{1}{8}L\rho^{2}a\Lambda+\frac{(11a^2-1)}{640a}L\rho^4\Lambda^2+...
\end{equation}

\subsubsection{Tolman prescription}
Using Table II and Eq.(\ref{metric car static =/ 0}), we find that
complicated quantities of $ U_{t}^{tx} $ and $ U_{t}^{ty} $  are
only non-zero needed components of super-potential. After surface
integration (Table II) and Taylor expansion around $\Lambda=0$ we
obtain:
\begin{equation}
_{T}E=\frac{(1-a^2)}{8a}L+\frac{(3a^2-1)}{16a}L\rho^{2}\Lambda-\frac{(17a^2-2)}{320a}L\rho^4\Lambda^2+...
\end{equation}

\subsubsection{Weinberg prescription}
Using Table II and Eq.(\ref{metric car static =/ 0}), we find that
complicated quantities of $ D^{xtt} $ and $ D^{ytt} $  are only
non-zero needed components of super-potential. After surface
integration (Table II) and Taylor expansion around $\Lambda=0$ we
obtain:
\begin{equation}
_{W}E=\frac{(1-a^2)}{4a^2}L+\frac{1}{4}L\rho^{2}\Lambda
+\frac{(34a^4-7a^2-2)}{160a^4}L\rho^4\Lambda^2+...
\end{equation}

\subsection{Non-static Cosmic String}
With $ \alpha=2\sqrt{\frac{\Lambda}{3}} $  and  $ a=1-4G\mu $
different energy-momentum prescriptions can be evaluated as
follows.

\subsubsection{M\o ller prescription}
Using Table II and Eq.(\ref{metric car non static}) we find that
non-zero components of $ \chi_{i}^{kl} $  are:
\begin{equation}
\chi_{x}^{x y}=-\chi_{x}^{y x}=\frac{(a-1)e^{\frac{1}{2}\alpha
t}}{a(x^2+y^2)}y
\end{equation}
\begin{equation}
\chi_{y}^{y x}=-\chi_{y}^{x y}=\frac{(a-1)e^{\frac{1}{2}\alpha
t}}{a(x^2+y^2)}x
\end{equation}

After surface integration, we find that integral of energy
vanishes. As we expect, in polar cylindrical coordinate system we
obtain same result for the energy integral as we have obtained in
cartesian coordinate system i.e. $_{M\o ller}E=0$.

\subsubsection{Landau-Lifshitz prescription}
Using Table II and Eq.(\ref{metric car non static}) we find that
non-zero needed components of $ \lambda^{iklm} $ are:
\begin{equation}
\lambda^{ttxx}=-e^{2\alpha t}\frac{y^2+a^2x^2}{x^2+y^2}
\end{equation}
\begin{equation}
\lambda^{ttyy}=-e^{2\alpha t}\frac{x^2+a^2y^2}{x^2+y^2}
\end{equation}
\begin{equation}
\lambda^{ttzz}=-e^{2\alpha t}a^2
\end{equation}
\begin{equation}
\lambda^{ttxy}=\lambda^{ttyx}=e^{2\alpha
t}\frac{(1-a^2)}{x^2+y^2}xy
\end{equation}
After surface integration (Table II) we obtain:
\begin{equation}
_{LL}E=\frac{(1-a^2)}{8}e^{2\alpha t}L
\end{equation}

\subsubsection{Papapetrou prescription}
Using Table II and Eq.(\ref{metric car non static}) we find that
non-zero needed components of $ N^{iklm} $ are:
\begin{equation}
N^{ttxx}=-\frac{(y^2+a^2x^2)+e^{\alpha
t}a^2(x^2+y^2)}{a(x^2+y^2)}e^{\frac{1}{2}\alpha t}
\end{equation}
\begin{equation}
N^{ttyy}=-\frac{(y^2+a^2y^2)+e^{\alpha
t}a^2(x^2+y^2)}{a(x^2+y^2)}e^{\frac{1}{2}\alpha t}
\end{equation}
\begin{equation}
N^{ttzz}=-(1+e^{\alpha t})a e^{\frac{1}{2}\alpha t}
\end{equation}
\begin{equation}
\frac{(1-a^2)}{a(x^2+y^2)}xy e^{\frac{1}{2}\alpha t}
\end{equation}
After surface integration (Table II) we obtain:
\begin{equation}
_{P}E=\frac{(1-a^2)}{8a}e^{\frac{1}{2}\alpha t}L
\end{equation}

\subsubsection{Einstein prescription}
Using Table II and Eq.(\ref{metric car non static}) we obtain
non-zero needed components of $ H_{i}^{kl} $ :
\begin{equation}
H_{t}^{tx}=\frac{(1-a^2)e^{\frac{1}{2}}}{a(x^2+y^2)}x
\end{equation}
\begin{equation}
H_{t}^{ty}=\frac{(1-a^2)e^{\frac{1}{2}}}{a(x^2+y^2)}y
\end{equation}
After surface integration (Table II) we obtain:
\begin{equation}
_{E}E=\frac{(1-a^2)}{8a}e^{\frac{1}{2}\alpha t}L
\end{equation}

\subsubsection{Bergmann prescription}
Using Table II,Eq.(33) and Eq.(\ref{metric car non static}) we find
non-zero needed components of $ B^{ikl} $ as:
\begin{equation}
B^{ttx}=\frac{(1-a^2)e^{\frac{1}{2}}}{a(x^2+y^2)}x
\end{equation}
\begin{equation}
B^{tty}=\frac{(1-a^2)e^{\frac{1}{2}}}{a(x^2+y^2)}y
\end{equation}
After surface integration (Table II) we obtain:
\begin{equation}
_{B}E=\frac{(1-a^2)}{8a}e^{\frac{1}{2}\alpha t}L
\end{equation}

\subsubsection{Tolman prescription}
Using Table II and Eq.(\ref{metric car non static}) we find that
non-zero needed components of $ U_{i}^{kl} $ are:
\begin{equation}
U_{t}^{tx}=\frac{(1-a^2)e^{\frac{1}{2}}}{2a(x^2+y^2)}x
\end{equation}
\begin{equation}
U_{t}^{ty}=\frac{(1-a^2)e^{\frac{1}{2}}}{2a(x^2+y^2)}y
\end{equation}
After surface integration (Table II) we obtain:
\begin{equation}
_{T}E=\frac{(1-a^2)}{8a}e^{\frac{1}{2}\alpha t}L
\end{equation}

\subsubsection{Weinberg prescription}
Using Table II and Eq.(\ref{metric car non static}) we obtain that
non-zero needed components of $ D^{ikl} $ are:
\begin{equation}
D^{xtt}=\frac{(1-a^2)(e^{\alpha
t}a^2-1-a^2)}{a^4(x^2+y^2)}e^{-2\alpha t}x
\end{equation}
\begin{equation}
D^{ytt}=\frac{(1-a^2)(e^{\alpha
t}a^2-1-a^2)}{a^4(x^2+y^2)}e^{-2\alpha t}y
\end{equation}
After surface integration (Table II) we obtain:
\begin{equation}
_{W}E=\frac{(1-a^2)(-e^{\alpha t}a^2+1+a^2)}{4a^2}e^{-2\alpha t}L
\end{equation}

\section{Conclusions and Remarks}
In previous section we calculated the energy of static and
non-static cosmic string space-times in a cylinder with length $L$
and radius $\rho$ surrounding the string symmetrically. We have
summarized all obtained results in table III. This work is one of a series of studies by the authors on energy-momentum prescriptions in general relativity~\cite{MA1, MA3, MA4}. 

\begin{table*}
\centering \caption{Energy of three different cosmic string
space-times in a cylinder with length $L$ and radius $\rho$
surrounding symmetrically the string ( $
\alpha=2\sqrt{\frac{\Lambda}{3}}$ and $ a=1-4G\mu $ )}
\label{table2}
\begin{tabular}{|c|c|c|c|}
\hline  &&&\\
Prescriptions& Static& Static  & Non-static\\
&($\Lambda=0$)&($\Lambda\neq0$)& ($\Lambda\neq0$)\\
\hline &&&\\

M\o ller & 0 & $-\frac{1}{4}a\rho^{2}L\Lambda+\frac{1}{16}a\rho^{4}L\Lambda^{2}-\frac{1}{160}a\rho^{6}L\Lambda^{3}+...$ & 0 \\
&&&\\\hline &&&\\

Landau-Lifshitz & $\frac{L(1-a^2)}{8}$ & $\frac{(1-a^2)}{8}L+\frac{(3a^2-1)}{16}L\rho^{2}\Lambda-\frac{(7a^2-1)}{128}L\rho^4\Lambda^2+...$ & $\frac{(1-a^2)}{8}e^{2\alpha t}L$ \\
&&&\\\hline &&&\\

Papapetrou & $\frac{L(1-a^2)}{8a}$&$\frac{(1-a^2)}{8a}L+ \frac{(3a^2-1)}{16a}L \rho^2 \Lambda- \frac{(17a^2-2)}{320a}L \rho^4 \Lambda^2+ ...  $ & $\frac{(1-a^2)}{8a}e^{\frac{1}{2}\alpha t}L$ \\
&&&\\\hline &&&\\

Einstein & $\frac{L(1-a^2)}{8a}$& $\frac{(1-a^2)}{8a}L+ \frac{(3a^2-1)}{16a}L \rho^2 \Lambda- \frac{(17a^2-2)}{320a}L \rho^4 \Lambda^2+ ...  $ & $\frac{(1-a^2)}{8a}e^{\frac{1}{2}\alpha t}L$ \\
&&&\\\hline &&&\\

Bergmann & $\frac{L(1-a^2)}{8a}$& $\frac{(1-a^2)}{8a}L+ \frac{a}{8}L \rho^2 \Lambda+ \frac{1}{640}\frac{(11a^2-1)}{a}L \rho^4 \Lambda^2+ ...$ & $\frac{(1-a^2)}{8a}e^{\frac{1}{2}\alpha t}L$\\
&&&\\\hline &&&\\

Tolman & $\frac{L(1-a^2)}{8a}$& $\frac{(1-a^2)}{8a}L+ \frac{(3a^2-1)}{16a}L \rho^2 \Lambda- \frac{(17a^2-2)}{320a}L \rho^4 \Lambda^2+ ...  $ & $\frac{(1-a^2)}{8a}e^{\frac{1}{2}\alpha t}L$ \\
&&&\\\hline &&&\\

Weinberg& $\frac{L(1-a^2)}{4a^2}$ & $\frac{(1-a^2)}{4a^2}L+\frac{1}{4}L\rho^{2}\Lambda+\frac {(34a^4-7a^2-2)}{160a^4}L\rho^4\Lambda^2+... $ & $\frac{(1-a^2)}{4a^2}(-e^{\alpha t}a^2+1+a^2)e^{-2\alpha t}L $\\
&&&\\\hline

\end{tabular}
\end{table*}

Regarding contents of table III:
\begin{itemize}
 \item It is concluded that the energy is turn out to be finite and well-defined in all these prescriptions for these space-times.
 \item Substituting $a=1$ in first column all prescriptions give energy equal to zero that is completely consistent. Because, if $a=1$, Eq.(\ref{metric static=0}) reduces to the Minkowski line element which its energy is equal to zero in any arbitrary  region.
 \item For static ($\Lambda\neq0$) cosmic string space-time Einstein, Tolman, and Papapetrou's prescriptions lead to the same results. In addition, when $\Lambda\rightarrow 0$, Bergmann prescription joins to this list. For the non-static case Einstein, Papapetrou, Tolman, and Bergmann prescriptions have the same result. This coincidence supports and extends Virbhadra's viewpoint [26] that different energy-momentum prescriptions may provide some basis to define a unique quantity. However, the remaining prescriptions give different energy densities (because of non-covariant property of pseudo-tensors).
 \item As we expect, for $\Lambda\rightarrow 0$ energy expressions in the second and third columns reduce to their corresponding expressions in the first column. It should be noticed that we calculated the components of the second column separately, by using line element Eq.(\ref{metric static=0}) in energy-momentum prescriptions.
 \item Unlike other prescriptions, M\o ller's prescription leads to a zero quantity for energy. This shortcoming is in contradiction with Lessner's viewpoint and supports Virbhadra's conclusion. Lessner [14] believed that M\o ller's prescription is a powerful tool for calculating the energy-momentum pseudo-tensors in GR, and Virbhadra [13] concluded that Einstein's energy-momentum prescription is still the best available method for computing energy-momentum in a given space-time.
 \item Reviewing the results shows that adding a factor $a$ in denominators of Landau-lifshitz's results causes that this prescription also give equivalent results (in comparison with Einstein, Tolman, and Papapetrou's prescriptions). In other words Landau-Lifshetz's results are different with other similar results (Einstein, Tolman, and Papapetrou) just in a factor $a$ in denominator. This dilemma is due to the fact that the conserved quantity in Landau-Lifshitz prescription is $ _{\textrm{\tiny eff}}T_{a}^{b}=(-g)(T_{a}^{b}+t_{a}^{b}) $ (weight +2)  instead of $ _{\textrm{\tiny eff}}T_{a}^{b}=\sqrt{-g}(T_{a}^{b}+t_{a}^{b}) $  (see Eq.(3) and Reference [4]) which its weight is +1. So, we should be careful about using this expression with weight +2 in our integration (see [33] chapter 7). Calculating energy with using a correction to Landau-lifshitz prescription i.e.  $\lambda^{iklm}=\sqrt{-g}(g^{ik}g^{lm}-g^{il}g^{km})_{,m}$  instead of $\lambda^{iklm}=(-g)(g^{ik}g^{lm}-g^{i
 l} g^{km})_{,m}$ (Table II) leads to consistent results.
\item In the final remark we would like to raise some points on the validity of the metric (\ref{metric static =/ 0}). Eq.(\ref{metric static =/ 0})
with $\mu \rightarrow 0$ faces with some problems to present the deSitter space-time (\textbf{dSS}). It has intrinsic singularities at $\rho=\frac{n\pi}{\sqrt{3\Lambda}}\;,\;\;\;n=odd$ \cite{BL}, while \textbf{dSS}
is free of them. The standard form of \textbf{dss} is:
\begin{eqnarray}\label{A}
ds^2 = dt^2 &-&\frac 3\Lambda\cosh ^2(\sqrt{\frac 3\Lambda}t)\times\nonumber\\
&\;&(d\chi^2+\sin ^2\chi(d\theta ^2+\sin ^2\theta d\phi ^2))
\end{eqnarray}
where $-\infty< t<+\infty,0\leq\chi\leq \pi, 0\leq\theta\leq\pi,\;0\leq\phi\leq 2\pi$.
By the following transformations:
\begin{eqnarray}
\hat{t}&=&\sqrt{\frac 3\Lambda}\log(\sinh(\sqrt{\frac 3\Lambda}t)+\cosh(\sqrt{\frac 3\Lambda}t)+\cos\chi)\\
\hat{x}&=&\frac{\cosh(\sqrt{\frac 3\Lambda}t)\sin\chi\cos\theta}{\sinh(\sqrt{\frac 3\Lambda}t)+\cosh(\sqrt{\frac 3\Lambda}t)+\cos\chi}\\
\hat{y}&=&\frac{\cosh(\sqrt{\frac 3\Lambda}t)\sin\theta\cos\phi}{\sinh(\sqrt{\frac 3\Lambda}t)+\cosh(\sqrt{\frac 3\Lambda}t)+\cos\chi}\\
\hat{z}&=&\frac{\cosh(\sqrt{\frac 3\Lambda}t)\sin\theta\sin\phi}{\sinh(\sqrt{\frac 3\Lambda}t)+\cosh(\sqrt{\frac 3\Lambda}t)+\cos\chi}
\end{eqnarray}
The metric (\ref{A}) transforms to the steady state form or the so-called half deSitter metric:
\begin{equation}
ds^2=d\hat{t}^2-\exp(2\sqrt{\frac 3\Lambda}t)
(d\hat{x}^2+d\hat{y}^2+d\hat{z}^2).
\end{equation}
In polar coordinate systems it takes the form:
\begin{eqnarray}\label{B}
ds^2=d\hat{t}^2&-&\exp(2\sqrt{\frac 3\Lambda}t)\times\nonumber\\&\;&(d\hat{r}^2+\hat{r}^2(d\theta ^2+\sin ^2\theta d\phi ^2))
\end{eqnarray}
Then by transformation:
\begin{eqnarray}
&\;&r=exp(\sqrt{\frac 3\Lambda}t)\hat{r}\\
&\;&t=\hat{t}-\frac12\sqrt{\frac 3\Lambda}\ln(1-\frac \Lambda 3 r^2)
\end{eqnarray}
The metric (\ref{B}) can be transformed to the static form:
\begin{eqnarray}\label{C}
ds^2=(1-\frac \Lambda 3 r^2)dt^2&-&\frac 1{(1-\frac \Lambda 3 r^2)} d r^2\nonumber\\
&-&r^2(d\theta^2+\sin^2\theta d\phi^2)
\end{eqnarray}
Writing the metric (\ref{C}) in cylindrical coordinate systems $(t,\rho,\varphi,z)$ we have:
\begin{eqnarray}\label{D}
ds^2&=&(1-\frac \Lambda 3 r^2)dt^2-\frac{1-\frac \Lambda3 z^2}{1-\frac \Lambda 3 r^2}d\rho^2\nonumber\\
&-&2\frac{\frac \Lambda3 \rho z}{1-\frac \Lambda 3 r^2}d\rho dz-\frac{1-\frac \Lambda3 \rho^2}{1-\frac \Lambda 3 r^2}dz^2\nonumber\\&-&\rho^2d\varphi^2
\end{eqnarray}
Now the difference between Eq.(\ref{D}) and  Eq.(\ref{metric static =/ 0}) when $\mu\rightarrow 0$ is quite evident. This means although the 
static metric (\ref{metric static =/ 0}) is an exact solution for Einstein equations
but there is some doubt as to whether it actually fulfills precisely the requirements of the space-time associated with a cylindrical cosmic string
located in a cosmological constant background.
\end{itemize}
\acknowledgments{A.M.A. wants to appreciate supports of research council at University of Tehran.}

\end{document}